# PERFORMANCE STUDY OF TIME SERIES DATABASES

Bonil Shah[1], P. M. Jat[2] and Kalyan Sasidhar[3]

[1]DAIICT, Gandhinagar, India
202011043@daiict.ac.in
[2]DAIICT, Gandhinagar, India
pm_jat@daiict.ac.in
[3]DAIICT, Gandhinagar, India
kalyan_sasidhar@daiict.ac.in

## ABSTRACT

*The growth of big-data sectors such as the Internet of Things (IoT) generates enormous volumes of data. As IoT devices generate a vast volume of time-series data, the Time Series Database (TSDB) popularity has grown alongside the rise of IoT. Time series databases are developed to manage and analyze huge amounts of time series data. However, it is not easy to choose the best one from them. The most popular benchmarks compare the performance of different databases to each other but use random or synthetic data that applies to only one domain. As a result, these benchmarks may not always accurately represent real-world performance. It is required to comprehensively compare the performance of time series databases with real datasets. The experiment shows significant performance differences for data injection time and query execution time when comparing real and synthetic datasets. The results are reported and analyzed.*

## 1. INTRODUCTION

The importance of sensors has been growing in recent years as the fourth industrial revolution has begun. The increased use of sensors results in a rise in the volume of sensor data that must be stored and processed. The Database Management System must be scaled appropriately to handle that volume of data. There are differing opinions on whether typical Relational Database Management Systems can handle such a large volume of data. Knowing that sensor data has a timestamp associated with it most of the time and can thus be utilized as an index, a trade-off between RDBMS and NoSQL DBMS arose: Time Series Databases.

Time series data are utilized in the finance business (e.g., historical stock performance research), the IoT industry (e.g., capturing the reading from sensors), and the analytics industry (e.g., tracking performance over time). Time series data is used for sensors, networks, the stock market, and other applications[4]. Time series databases make it easy to manage events that are tracked, monitored, downsampled, and aggregated over time[12, 14]. A few years ago, Time Series Databases were not so popular. It was only used by some trading applications or any other applications where you required data monitoring. Everyone else was using relational databases or NoSQL back then. But in the last three-four years, TSDBs like InfluxDB, OpenTSDB, and TimescaleDB have gained popularity.

There are numerous time series benchmarks available, which are used to compare database performance, such as IoTBenchmark, TSDBBench, TS-Benchmark, etc. However, all of these benchmarks measure the performance of time series databases using random or synthetic datasets.

The biggest issue with using pre-captured or pre-calculated data is that it only applies to one circumstance. The queries in these benchmarks are also based on the datasets. It is required to study how the time series databases perform on real datasets of different domains like finance, analytics, and IoT. It is also important to consider various types of queries, such as queries with aggregate functions, exact point queries, queries with time range filters, etc. These are frequently used queries in industries to access historical data.

## 2. LITERATURE REVIEW

This section represents the literature survey and is structured as follows: Section 2.1 summarises all the types of databases, like relational and non-relational databases. Section 2.2 presents detailed information about time series databases and their use cases. Section 2.3 has information about some existing time series benchmarks.

### 2.1. Databases

A database is a structured collection of data that can be accessed and managed efficiently. Databases are used to store, maintain, and access any type of data. They gather data on people, places, or things. That data is collected in one location so that it can be observed and analyzed.

#### 2.1.1. Database Management System

A database management system (or DBMS) is the application used to manage databases. The system provides users with the ability to perform various operations such as create, read, write, and delete. MySQL, Oracle, etc., are some popular examples of DBMS used in different applications.

#### 2.1.2. Relational Databases

Relational databases are based on the relational model, a simple and obvious manner of representing data in tables. Each row in the table is a record with a unique ID known as the key in a relational database. The table's columns carry data attributes, and each record typically includes a value for each attribute, making it simple to construct links between data points. SQL is the query language used to manipulate and retrieve the data from the relational database.

#### 2.1.3. Non-Relational Databases

Non-relational databases are often known as "NoSQL," which stands for Not Only SQL. It is a database that does not use the tabular schema of rows and columns that is common in most traditional database systems. Non-relational databases, on the other hand, use a storage model that is optimized for the specific needs of the data being stored. Following are some characteristics of non-relational databases: The scalability of the non-relational database is very high, which makes it suited for massive amounts of data. NoSQL databases, in comparison to SQL databases, are not complex. They store data in an unstructured or semi-structured format that does not require any relational or tabular structure. NoSQL databases are highly durable as they can accommodate data ranging from heterogeneous to homogeneous.

### 2.2. Time Series Databases

Time series databases are used to manage time series data. A time series is an ordered sequence of variable values over a specified time interval. Time Series is a popular format for representing and analyzing data that changes over time[13]. Formally, a time series is a description of a stochastic process consisting of a collection of pairs $[(p_1, t_1), (p_2, t_2), ... (p_n, t_n)]$ where $p_i$ is the information collected in time $t_i$. A DBMS that can (i) store a record consisting of a timestamp, value, and optional tags, (ii) store multiple records aggregated, (iii) query for records, and (iv) contain time ranges in a query is called time series databases.

### 2.2.1. Properties

S. N. Z. Naqvi, S. Yfantidou, and E. Zimányi have mentioned some characteristics of time series data in their paper called "Time Series Database and InfluxDB"[2]. Following are some characteristics of time series data:

• **Scalability:** The volume of time-series data is rapidly increasing. Every hour, for example, the linked car will send 25 GB of data to the cloud. Regular databases are not built to handle this level of scalability. On the other hand, Time series databases are designed to account for scale by introducing functions that are only possible when time is prioritized. This can improve performance, such as higher insertion rates, faster queries at scale, and better data compression.

• **Usability:** TSDBs often incorporate time series data analysis capabilities and processes. They use data retention policies, continuous queries, flexible temporal aggregations, range queries, etc. As a result, this improves usability by improving the user experience while dealing with time-related analysis.

• **Data Compression:** Time Series data is mostly recorded per second or even less time, so there will be a lot of data to handle. A good data compression technique is required to handle this kind of data.

• **Data Location:** The queries on data are slow when the related data is not together. Time series databases co-relate the chunks of data of the same device and time range to access the data more efficiently.

• **Fast and Easy Range Queries:** As data location is there in TSDB, the range queries are very fast compared to traditional databases. The query language in TSDB is also like SQL, so it is easier for users to write queries.

• **High Write Performance:** In peak load, most databases cannot quickly serve the read and write request. TSDBs should ensure high availability and performance for both read and write operations during peak loads because they are usually designed to stay available in these conditions.

### 2.2.2. Uses and Benefits

• **Time Series Analysis:** Time series analysis means to analyze the variable which changes over time. Ex. The temperature of different cities.

• **Time Series Forecasting:** Time Series Forecasting means predicting future activity by using information regarding associated patterns and historical values. For example, Weather Forecasting, Earthquake Forecasting, Sales Forecasting, etc.

• **Regression Analysis:** Regression analysis means how the changes of any one variable affect another variable, like in the stock market, how one stock price changes affect any other stock.

### 2.2.3. Industry Use Cases

The Internet of Things revolution has enabled gadgets that were previously only available as offline systems to be connected to the internet. These devices are classified into two types: actuators, which operate on commands, and sensors, which sense the current environment and convert physical quantities into digital values. The values transmitted by these sensors and the standard analyses performed on them are a natural fit for time series databases[4]. A timestamp is assigned to each data point generated by a sensor. The application domain determines the frequency at which data is generated. The intervals used by sensors are usually every minute, every 10 minutes, or every hour. Getting the most recent data points, averaging data points over time periods, and flexible visualization are all common operations on sensor data. IoT data sets are typically regular and low in volume for small numbers of sensors.

Time series can be used in the analytics domain to track website visits, advertisement clicks, or E-commerce orders. The data volume may be affected by a variety of factors, such as time (e.g., orders on Tuesday versus orders on Sunday), weather (e.g., the number of air conditioners sold in a specific store), or other arbitrary circumstances (e.g., the number of cars per hour on a day with a taxi strike).

In the financial domain, time series data is commonly used. Time series can be used to illustrate stock prices, currency rates, and portfolio valuations. As a result, storing financial data points in a time series database is a sensible choice. For example, Kx Systems' kdb+, a time series database, is frequently used in high-frequency trading. Other financial use cases specifically presented by kdb+ include algorithmic trading, FX trading, and regulatory management. Financial time series are regular, although the volume varies dramatically. Data points can be generated every day, every minute (for example, stock closing prices), or every few seconds. (e.g., high-frequency trading).

Physical systems are frequently monitored using time series databases. Most time series databases feature spatial data storage and querying capabilities. It is possible to link location data in this manner. Asset tracking (e.g., storing the current position of all taxis) and geographical filtering (e.g., a number of taxis within a range) are examples of use cases. Asset tracking use cases produce data points with geospatial information. Time series produced can be regular (e.g., location is sent every minute) but is often irregular.

**2.2.4. Some Databases and Their Designs**

TimescaleDB is an open-source time series database that is entirely based on PostgreSQL[15]. It inherits PostgreSQL's functionality and provides adaptive temporal chunks to reduce system resource usage. TimescaleDB offers full SQL support.

Druid is a columnar data storage that is open source and designed for OLAP queries[7]. It consists of many nodes that support a specific set of functionalities. Interactions between nodes have been reduced to a bare minimum.

InfluxDB is the most commonly used time series database. InfluxDB employs two protocols for data intake: a text-based protocol known as "Line Protocol" and a deprecated JSON protocol[9]. It uses its own query language, Flux. InfluxDB's storage engine is built on a time-structured merge tree (TSM) derived from a log-structured merge tree[13]. TSM significantly improves InfluxDB's performance when writing and reading data.

Cassandra is a NoSQL database that provides the always-on availability, fast read-write speed, and limitless linear scalability required by modern applications. Thousands of businesses trust Cassandra for its scalability and high availability without sacrificing performance. It is the ideal platform for mission-critical data because of its linear scalability and verified fault-tolerance on commodity hardware or cloud infrastructure[3].

**2.3. Time Series Benchmarks**

There are some existing time series benchmarks available in the market, which are examined in more detail in this section. The one common thing about these benchmarks is that they all use any random or generated dataset.

IoTDB Benchmark is built exclusively for TSDBs and IoT application scenarios. It focuses on specific data ingestion scenarios[11]. IotDB-benchmark has 10 types of queries, ranging from the exact point queries to time range queries with value filters. The performance metrics measured by the IoTDB benchmark are query latency, space consumption, system resource consumption, and throughput. The IoTDB benchmark's data generator generates random values within the provided range. Therefore, it is not using any real-world data. Many data generating factors, such as the data type of fields, the number of tags per device, and many other things, can be configured using the IoTDB-benchmark. The IoTDB benchmark currently supports IoTDB, InfluxDB, KariosDB, OpenTSDB, QuestDB, Sqlite, and TimescaleDB. This benchmark focuses solely on IoTDB, and not all functions are supported by databases other than IoTDB. For example, only IoTDB supports the generation and insertion of customized time series.

TSDBBench is a benchmark designed for comparing different TSDBs. In a project called YCSB-TS, it enhances the Yahoo! Cloud Serving Benchmark (YCSB) for use with time series databases[2]. In practice, the benchmark seems unmaintained. The documentation is outdated, and the required files are hosted on an inactive domain. The databases version used in this benchmark is also ancient. It only measures two metrics query latency and space consumption.

The TS-Benchmark is another TSDB benchmark centered on the requirements of managing huge time series data. It uses DCGAN based model to generate the high-quality synthesized data after being trained with real time series data[8]. For the benchmark, they used data from wind turbines. The number of wind farms, devices, and sensors utilized to generate data is viewed as a benchmark scale factor. TS-Benchmark can compare the performance of different TSDBs under various workloads such as data injection, data loading, and data fetching. It uses 6 queries with a different parameter to test the query latency of various TSDBs. The TS-Benchmark currently supports four types of TSDBs: InfluxDB, Druid, TimescaleDB, and OpenTSDB.

## 3. EXPERIMENTATION

Time series datasets can be categorized as synthetic data and real data. In the previous section, we have seen that all the benchmarks use synthetic data for comparison. It may represent the problem for the generalization of their results. This section describes a study focusing on real-world data. It also includes synthetic data to examine the performance differences. This section is structured as follows: Section 3.1 has the three performance metrics used to compare the databases. Section 3.2 represents the experimental setup of the experiment. Section 3.3 has details about the datasets we have used in the experiment. Section 3.4 represents all the types of quires used in the experiment.

### 3.1. Performance Metrics

Benchmarking requires defining what should be measured and how to evaluate the outcomes in order to create a ranking and distinguish between different results. A performance metric must be specified in order to do so. First, we used throughput to measure the data loading performance of individual databases. The time required to load the local dataset is called throughput.

Second, the space used to store a particular dataset is measured for every database. It will be interesting to see if there is any difference in space efficiency between TSDBs when storing the same data. If the space consumption of any database is low, it means that the database compresses the data at a higher rate.

Third, we used query latency to measure the time required to run the particular query. It measures how long a query takes to run from the time it is sent to the database until a result is delivered and received by the client. A query must be successfully and thoroughly completed. All eight types of queries will be executed and measured for each database.

### 3.2. Experimental Setup

When comparing different TSDBs, the results must be comparable if developed in the same environment. It is also essential that each TSDB executes the task under similar conditions, implying that the benchmark utilized does not give any TSDB an advantage or disadvantage. It is not expected that results from different situations (for example, measuring TSDB 1 on system A and TSDB 2 on system B) will be comparable. The databases are tested on Intel(R) Core(TM) i5-

9500 CPU @ 3.00GHz, 8GB of RAM, 500GB of hard disk, and the operating system of Ubuntu 20.04.1 64-bit.

InfluxDB v2.1, TimescaleDB v2.6, Druid v0.22.1, and Casandra v3.11.12 were evaluated and compared in this experiment. InfluxDB's cluster version is not free, and TimescaleDB has no cluster version. We considered single-node versions to compare them fairly.

### 3.3. Datasets

All the TSDB benchmarks available in the market are using only one particular dataset. In the experiment, four different datasets of different domains are considered. Of course, four distinct datasets are insufficient to cover every industry or use case. However, analyzing the results of benchmarks using these workload datasets will allow comparisons to show whether the examined use case has an impact on performance. Out of these four datasets, three are real datasets from different industry domains like Financial, IoT, and Analysis. We have also taken one synthetic dataset into consideration to compare the performance of synthetic data and real data. All the real datasets are taken from Kaggle[10, 6, 5], and for synthetic data, we used the timeseries-generator library in python to generate the random time series data. Timeseries-Generator package is an exciting and excellent way to generate time-series data. A generator is a linear function with several factors and a noise function in this case. It is the data of some fast food item sales in different countries.

### 3.4. Queries

In this experiment, we have considered total eight kinds of queries. There are almost all kinds of the queries like exact point queries, queries with aggregate functions(min, max, sum, avg, count), group by queries, and queries with a time range.

#### 3.4.1 Queries for Financial Dataset

1. Exact Point Query

- Get the price of some stock on some particular day

2. Aggregation Queries

- Get the total volume of a particular stock

- Get the minimum price of the particular stock

- Get the maximum price of the particular stock

- Count the number of times when the particular stock volume is greater than 5000

- Get the avg value of the stock price of a particular stock

3. Group By Query

- Get the maximum volume group by stock

4. Time Range Query

- Get the lowest and highest price of a particular stock from 2017 to 2019

#### 3.4.2 Queries for Analytics Dataset

1. Exact Point Query

- Get the amount paid on some particular taxi

2. Aggregation Queries

- Get the minimum distance

- Get the maximum distance
- Get total distance travel
- Get the total number of passengers who travels alone
- Get the average tip amount

3. Group By Query
- Get the total toll amount paid group by vendor

4. Time Range Query
- Get the total amounts from 1 Jan 2015 to 25 Jan 2015

### 3.4.3 Queries for IoT Dataset

1. Exact Point Query
- Get the turbidity on some particular day and time

2. Aggregation Queries
- Get the date and time when the temperature is lowest at Ohio street beach
- Get maximum turbidity at calumet beach
- Get the total turbidity of montrose beach
- Count when the temperature at calumet beach went below 19
- Get the average wave height of Montrose Beach

3. Group By Query
- Get the average battery life group by beach

4. Time Range Query
- Get top 50 days in 2018 when the temperature is highest

### 3.4.4 Queries for Synthetic Dataset

1. Exact Point Query
- Get the sale value of some particular day and time

2. Aggregation Queries
- Get the minimum sale value of pizza
- Get country name with maximum gdp factor
- Get the total value of sandwich
- Get the number of count when value is above 15000
- Get the average value of Netherlands

3. Group By Query
- Get the average value group by product

4. Time Range Query
- Get the value data from 2005 to 2015

# 4. RESULTS

This section represents the results of the experiment. It is structured as follows: Section 4.1 represents the results of data loading, Section 4.2 represents the results of space consumption, and Section 4.3 represents the results of queries.

## 4.1. Data Loading

The local data files are imported using the different database's built-in import tool. Before data import, the databases are restarted to ensure that no cache is used. Table 1 shows the data loading performance for every dataset.

Table 1. Data loading (in seconds).

| Database | IoT | Financial | Analytics | Synthetic |
|---|---|---|---|---|
| TimescaleDB | 8 | 949 | 604 | 44 |
| Druid | 6 | 695 | 303 | 106 |
| InfluxDB | 2 | 576 | 283 | 19 |
| Cassandra | 2.23 | 1068 | 388 | 24.993 |

Based on the experiment results, we can say that InfluxDB beats all of the datasets in terms of load performance. After InfluxDB, Druid is the best option. Druid requires permanent backup storage for the distributed file system, known as deep storage. Because each segment is repeated in deep storage, the time required to load the data increases. It outperforms TimescaleDB and Cassandra in real datasets, but it is the worst performer in synthetic datasets. So, here you can see the performance difference of Druid in real and synthetic data. Cassandra beats TimescaleDB in all datasets except the financial dataset. We can say from the results that InfluxDB is the best choice, followed by Druid, Cassandra, and TimescaleDB. If we compare the best and the worst performer, InfluxDB outperforms TimescaleDB in the IoT dataset by 4 times, the financial dataset by 1.5 times, the analysis dataset by nearly 3 times, and the synthetic dataset by almost 2 times.

## 4.2. Space Consumption

All four datasets are inserted in particular databases in CSV format. Afterward, the space occupied by the data for the specific database is measured. Table 2 shows the data loading performance for every dataset. In terms of space consumption, Druid outperforms all other databases by a wide margin. It took only 25\% of the original data, which is exceptional. It compresses data to a high degree using data granularity. The level of detail in a data structure is measured by data granularity. The granularity of measurement of time-series data, for example, could be based on intervals of years, months, weeks, days, or hours.

Table 2. Space consumption (in MB).

| Database | IoT | Financial | Analytics | Synthetic |
|---|---|---|---|---|
| CSV | 3.9 | 3844.83 | 1985.96 | 130.9 |
| TimescaleDB | 9.7 | 5222.4 | 2355.2 | 179.4 |
| Druid | 6.14 | 946.43 | 489.76 | 56.15 |
| InfluxDB | 3.6 | 3276.8 | 1536 | 126 |
| Cassandra | 3.9 | 4526.08 | 1433.6 | 94.51 |

InfluxDB is the next option after Druid. InfluxDB uses many compression techniques for various data types. It also takes up less space than the original file, but it is more than Druid. Cassandra also takes less space than the original in all the datasets. However, Cassandra takes up more space in the financial dataset than the original. So, as data size grows, Cassandra's performance decreases. In terms of space utilization, TimescaleDB is the worst performer. Because TimescaleDB is built on Postgres, storing data in tabular format takes much space. If we compare the best and the worst performer, Druid outperforms TimescaleDB in the IoT dataset by 1.5 times, the financial dataset by 5 times, the analysis dataset by nearly 3 times, and the synthetic dataset by almost 5 times.

### 4.3. Query Latency

The results of queries are divided into four groups. It is structured as follows: Section 4.3.1 represents exact point query, Section 4.3.2 represents aggregation queries, Section 4.3.3 represents group by query, and Section 4.3.4. represents time range query.

#### 4.3.1. Exact Point Query

When we performed the exact point query on all databases, TimescaleDB performed better than all other databases. This is because TimescaleDB is based on a relational database, PostgreSQL, and uses time as the primary key, which is specially indexed. The results are shown in Figure 1. In the financial domain, Druid beats TimescaleDB because that dataset is the largest. After TimescaleDB, Druid is the second-best option, followed by InfluxDB and Cassandra. TimescaleDB outperforms Cassandra in the IoT query workload by 3 times, the financial query workload by 14 times, the analysis query workload by nearly 9 times, and the synthetic query workload by nearly 4 times. Clearly, the query workload has a significant impact on performance.

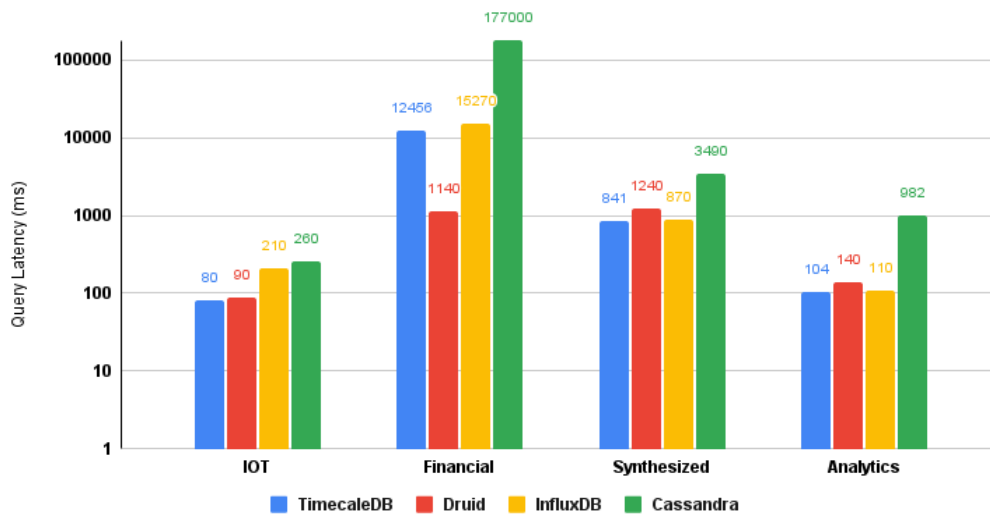

Figure 1. Exact point query

#### 4.3.2. Aggregation Queries

We have used the aggregation functions MIN, MAX, COUNT, SUM, and AVG in aggregate queries in the experiment. The results are pretty similar for all aggregate functions. The results of MIN, MAX, COUNT, SUM, and AVG are shown in Figure 2, 3, 4, 5, and 6, respectively. Except for the financial dataset, the results show that InfluxDB outperforms all other databases in every aggregate function query. InfluxDB beats the weakest performer, TimescaleDB, by around 300 times, which is impressive. In terms of query latency performance, Druid is pretty close to InfluxDB. Cassandra is a NoSQL database. However, it outperformed TimescaleDb, which is specifically designed to handle time series data. For the IoT query workload, InfluxDB performs 5 times better than TimescaleDB, for the financial query workload 56 times better, for the analysis query workload nearly 4 times better, and for synthetic data almost 18.5 times better.

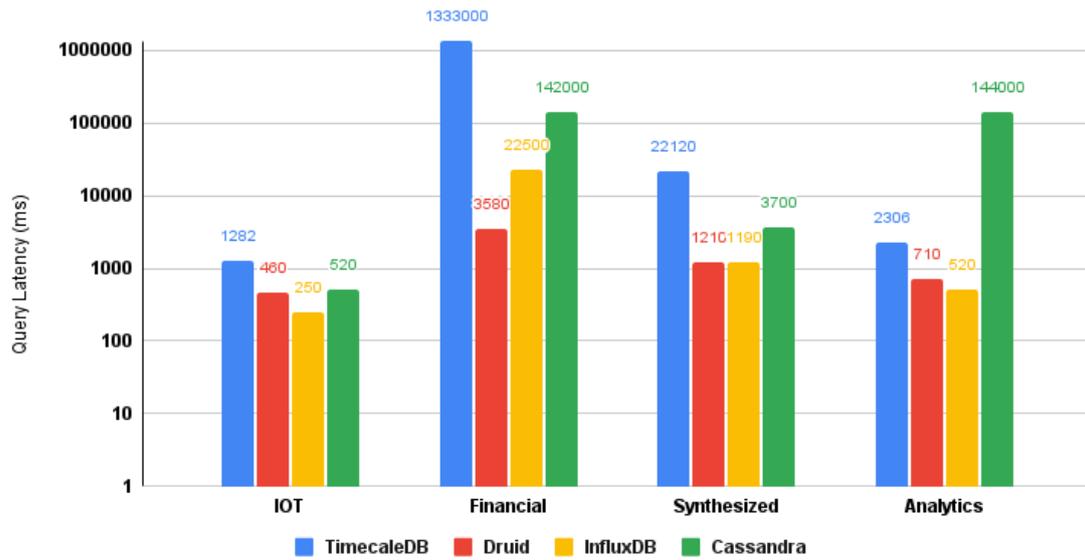

Figure 2. MIN Aggregation query

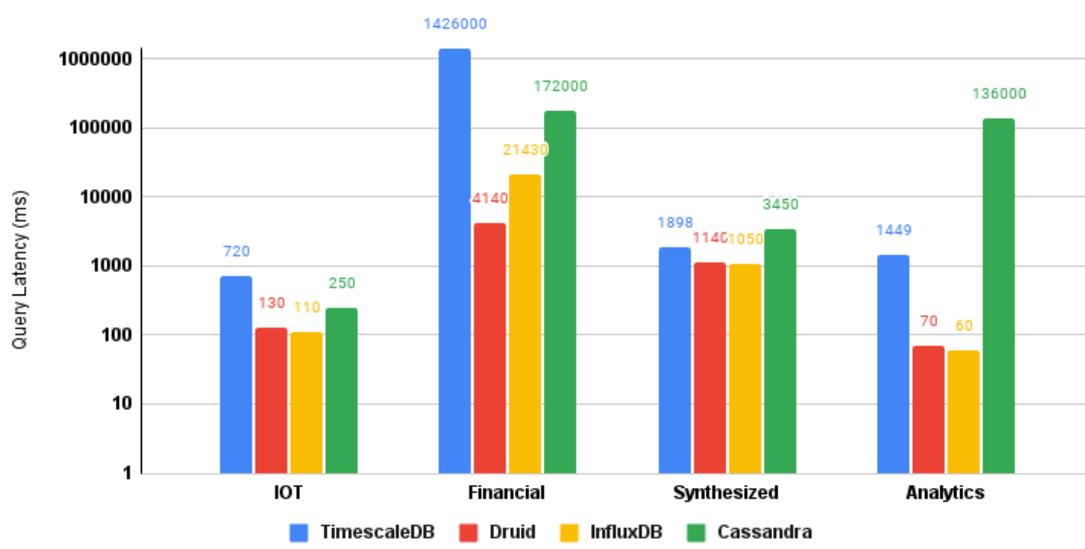

Figure 3. MAX Aggregation query

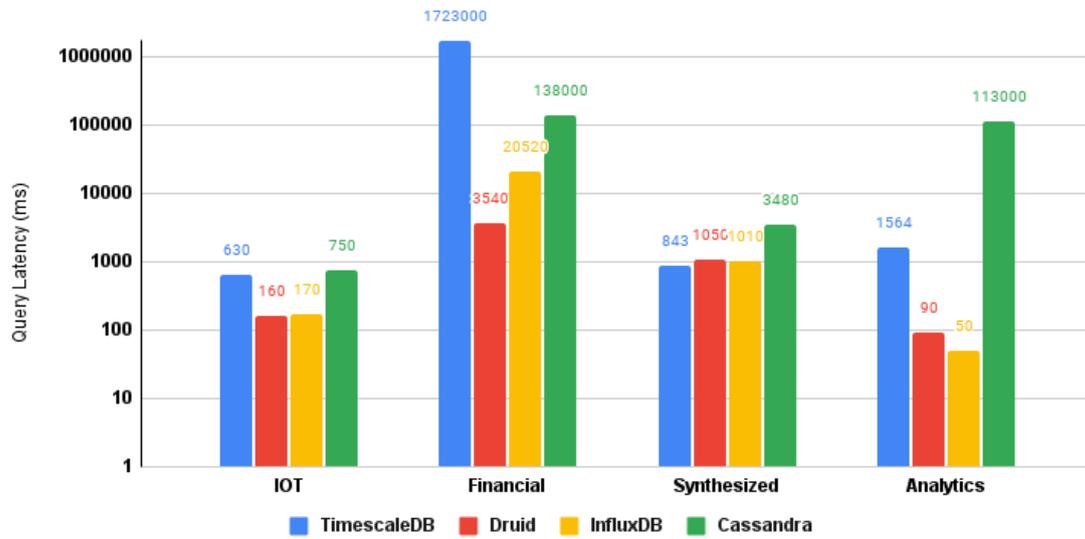

Figure 4. COUNT Aggregation query

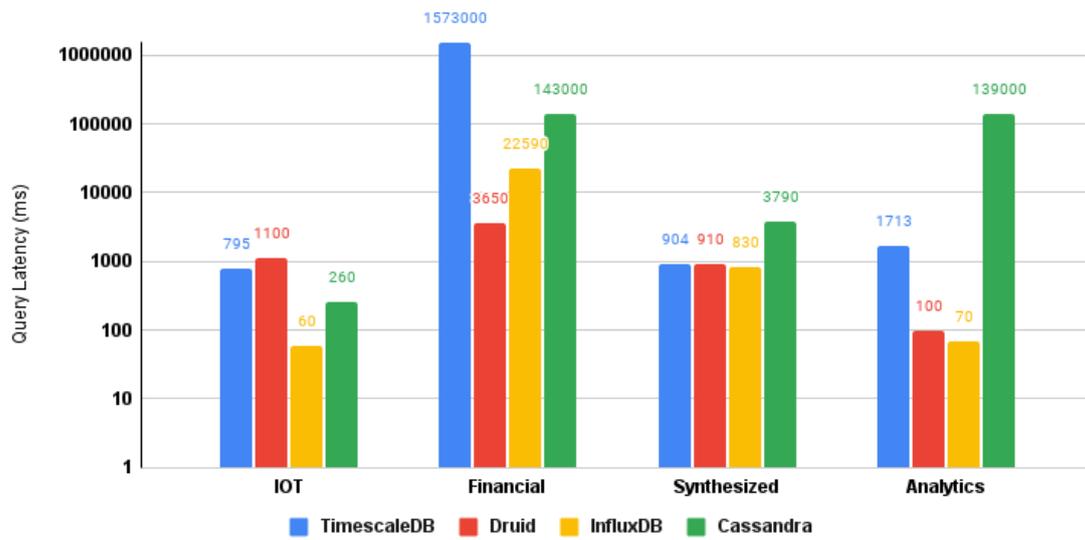

Figure 5. SUM Aggregation query

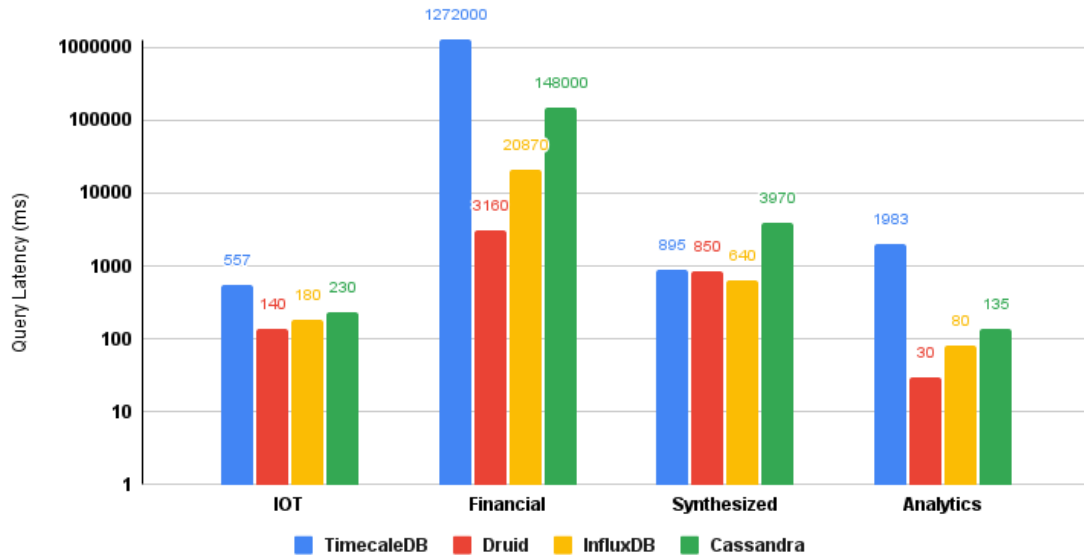

Figure 6. AVG Aggregation query

### 4.3.3. Group By Query

The most common query type in the industries is a group by queries. we ran group by queries on various datasets across all databases. The results are shown in Figure 7. Druid and InfluxDB performed equally well across all datasets in this query type. Druid outperforms InfluxDB by a hair, but it's a razor's edge. After Druid and InfluxDB, Cassandra outperforms TimescaleDB in the IoT and financial domains, whereas TimescaleDB outperforms Cassandra in the analytics domain.

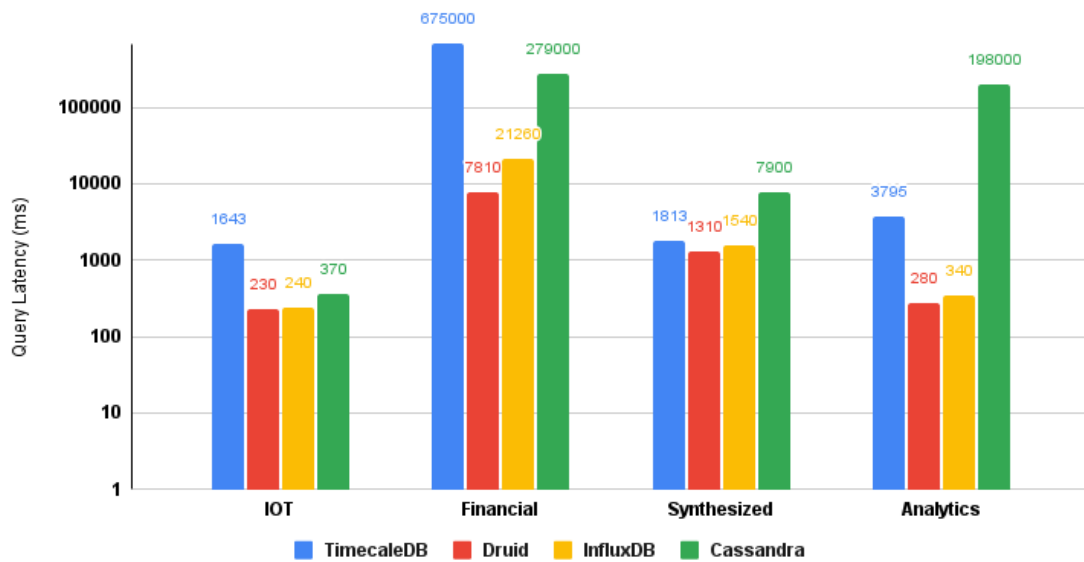

Figure 7. Group by query

### 4.3.4. Time Range Query

The time range query is the essential query type of time series data. It is used to retrieve data from a specific time range. The results of all the databases are shown in Figure 8. In the time range query workload, we can see that Druid outperforms all other databases. Druid is just ahead of

InfluxDB. In this query type also, Cassandra and TimescaleDb are the worst performers. Cassandra performs well in IoT and financial scenarios, and in other cases, TimescalDB performs better.

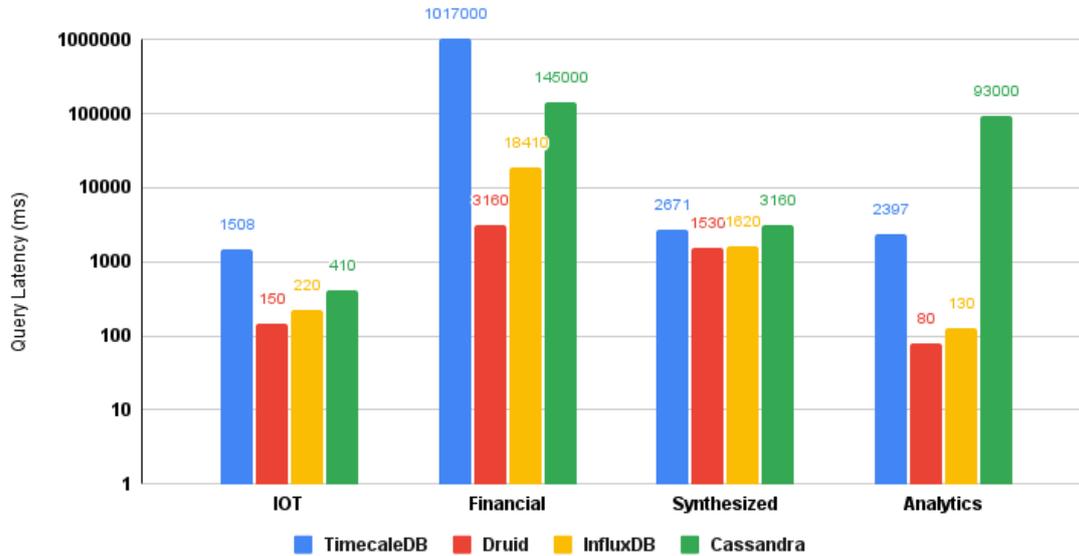

Figure 8. Time range query

## 5. CONCLUSION

In this experiment, the performance of TSDBs is evaluated by several real-world scenarios such as the IoT domain, the Financial domain, and the Analytics domain. We have also considered the synthesized data for comparison. We develop performance metrics that are more suited to evaluate the effectiveness of TSDBs. It includes data loading, database space usage, and query performance. We ran experiments and examined the findings on three common TSDBs, TimescaleDB, Druid, InfluxDB, and one NoSQL database, Cassandra. InfluxDB is a native TSDB with a storage engine that uses a Time-Structured Merge Tree (based on log-structured merge tree). TimescaleDB, based on PostgreSQL, uses row-wise storage to represent databases. Druid saves data in columnar format and creates indexes when data is injected. Cassandra is the most popular column-oriented NoSQL database.

From the experiment results, we have obtained some preliminary yet lightening conclusions: I) Because of its unique storage mechanism, namely a time-structured merge tree that supports high-speed writing, InfluxDB has a high injection and query throughput. II) During pre-processing, a granularity component in Druid separates data into segments and generates a bitmap index. Druid stream processing also struggles with excessive concurrency. Columnar storage format and indexes, on the other hand, pay off later. Druid achieves high query throughput, sometimes beating InfluxDB. III) Because TimescaleDB uses a row-wise storage format borrowed from PostgreSQL, it has poor query speed. The reason for low performance is that TimescaleDB does not automatically create indexes. IV) Cassandra is the NoSQL database that uses column format to store data. It gives better read and write performance than TimescaleDB.

We have also used one synthetic dataset with the other real datasets. The reason to use a synthetic dataset is to see if we can find any interesting results. There is a lot of difference between the synthesized and real datasets in data loading. Druid performs better in the real datasets, but it is the worst performer in synthesized data. The financial dataset also has different query performance and space consumption results than the other datasets. The recommendations are based on the performance comparison: I) If the focus lies on load performance, InfluxDB is the best choice, followed by Druid, Cassandra, and TimescaleDB. II) Druid is far better than others if the focus lies on the lowest space consumption. InfluxDB is the second-best choice. III) If there are more aggregate queries, one can use InfluxDB. It performs very well, followed by Druid and Cassandra. If there are queries containing data filters like time range, Druid is the best choice.

## REFERENCES

[1] Engines ranking. https://db-engines.com/en/ranking/time+series+dbms.

[2] A. Bader, O. Kopp, and M. Falkenthal. Survey and comparison of open source time series databases. Datenbanksysteme für Business, Technologie und Web (BTW 2017)-Workshopband, 2017

[3] Cassandra. Open source nosql database. https://cassandra.apache.org/_/index.html.

[4] S. Di Martino, L. Fiadone, A. Peron, A. Riccabone, and V. N. Vitale. Industrial internet of things: persistence for time series with nosql databases. In 2019 IEEE 28th International Conference on Enabling Technologies: Infrastructure for Collaborative Enterprises (WETICE), pages 340–345. IEEE, 2019.

[5] C. P. District. Beach water quality - automated sensors: City of chicago: Data portal. https://data.cityofchicago.org/Parks-Recreation/Beach-Water-Quality-Automated-Sensors/qmqz-2xku

[6] Elemento. Nyc yellow taxi trip data. https://www.kaggle.com/elemento/nyc-yellow-taxi-trip-data.

[7] A. S. Foundation. Database for modern analytics applications. https://druid.apache.org/

[8] Y. Hao, X. Qin, Y. Chen, Y. Li, X. Sun, Y. Tao, X. Zhang, and X. Du. Ts-benchmark: A benchmark for time series databases. In 2021 IEEE 37th International Conference on Data Engineering (ICDE), pages 588–599. IEEE, 2021.

[9] InfluxDB. Open source time series database. https://www.influxdata.com/.

[10] H. Kumar. Stock market india. https://www.kaggle.com/hk7797/

[11] R. Liu and J. Yuan. Benchmarking time series databases with iotdb-benchmark for iot scenarios. arXiv preprint arXiv:1901.08304, 2019.

[12] S. N. Z. Naqvi, S. Yfantidou, and E. Zimányi. Time series databases and influxdb. Studienarbeit, Université Libre de Bruxelles, 12, 2017.

[13] P. O'Neil, E. Cheng, D. Gawlick, and E. O'Neil. The log-structured merge-tree (lsm-tree). Acta Informatica, 33(4):351–385, 1996.

[14] D. Ted and F. Ellen. Time series databases: New ways to store and access data.

[15] TimescaleDB. Timeseries database for postgresql. https://docs.timescale.com/